\documentclass[aps,prl,superscriptaddress,amsmath,amssymb,twocolumn]{revtex4}
\usepackage{graphicx}
\usepackage{hyperref}
\usepackage{color}
\definecolor{darkblue}{rgb}{0.0,0.0,0.5}
\hypersetup{colorlinks,breaklinks,
            linkcolor=darkblue,urlcolor=darkblue,
           anchorcolor=darkblue,citecolor=darkblue}

\date{\today}

\allowdisplaybreaks

\usepackage{sansmath}

\begin{document}

\title{Direct Calculation of Ice Homogeneous Nucleation Rate for a Molecular Model of Water}

\author{Amir Haji-Akbari}
\affiliation{Department of Chemical and Biological Engineering, Princeton University, Princeton, NJ 08540}

\author{Pablo G. Debenedetti}
\email{pdebene@exchange.princeton.edu}
\affiliation{Department of Chemical and Biological Engineering, Princeton University, Princeton, NJ 08540}

\begin{abstract}
{Ice formation is ubiquitous in nature, with important consequences in a variety of environments, including biological cells, soil, aircraft, transportation infrastructure and atmospheric clouds. However, its intrinsic kinetics and microscopic mechanism are  difficult to discern with current experiments. Molecular simulations of ice nucleation are also challenging, and direct rate calculations have only been performed for coarse-grained models of wate. For molecular models, only indirect estimates have been obtained, e.g.~by assuming the validity of classical nucleation theory. We use a path sampling approach to perform the first direct rate calculation of homogeneous nucleation of ice in a molecular model of water. We use TIP4P/Ice, the most accurate among  existing molecular models for studying ice polymorphs. By using a novel topological approach to distinguish different polymorphs, we are able to identify a freezing mechanism that involves a competition between cubic and hexagonal ice in the early stages of nucleation. In this competition, the cubic polymorph takes over since the addition of new topological structural motifs consistent with cubic ice leads to the formation of more compact crystallites. This is not true for topological hexagonal motifs, which give rise to elongated crystallites that are not able to grow.  This leads to transition states that are rich in cubic ice, and not the thermodynamically stable hexagonal polymorph. This mechanism provides a molecular explanation to the earlier experimental and computational observations of the preference for cubic ice in the literature. }
\end{abstract}

\keywords{nucleation | ice | molecular simulations | statistical mechanics}
\maketitle

Ice nucleation affects the behaviour of many systems~\cite{PadayacheeSF2009, ChamberlainEngGeol1979, PotapczukJAE2013, YeCJCE2013, BakerScience1997, KirkbyScience2002}. For example, the formation of ice crystals inside the cytoplasm can damage living cells~\cite{PadayacheeSF2009}. The amount of ice in a cloud determines both its light-absorbing properties~\cite{BakerScience1997} and its precipitation propensity~\cite{KirkbyScience2002}, and is therefore an important input parameter in many meteorological models~\cite{FowlerJClimate1996, MurrayGRL2015}. Yet, current experiments are incapable of uncovering the kinetics and the molecular mechanism of freezing due to their limited spatiotemporal resolution. The ice that nucleates homogeneously in the atmosphere and vapor chamber experiments is predominantly comprised of the cubic-rich stacking-disordered polymorph, not the thermodynamically stable hexagonal polymorph~\cite{MurrayNature2005, SalzmannPNAS2012}. This observation has been rationalized invoking the Ostwald step rule~\cite{MalkinPCCP2015}. However, the molecular origin of this preference is unknown due to the limited spatiotemporal resolution of  existing experimental techniques. Furthermore, experimental measurements of nucleation rates are only practical over narrow ranges of temperatures~\cite{TaborekPRB1985}, with any extrapolation being prone to large uncertainties.

Computer simulations are attractive alternatives in this quest, as they make it possible to obtain at any given thermodynamic condition a statistically representative sample of nucleation events that can then be used to estimate the rates and identify the mechanism of nucleation. This, however, has only been achieved~\cite{GalliPCCP2011, GalliNatComm2013, HajiAkbariFilmMolinero2014} for coarse-grained representations of water, such as mW~\cite{MolineroJPCB2009}. For the more realistic molecular force-fields, all the existing studies have relied either on launching a few $\mu$s-long molecular dynamics (MD) trajectories\cite{SanzJACS2013, Matsumoto2002}, or on applying external fields~\cite{KusalikPRL1994}, or biasing potentials along  pre-chosen reaction coordinates~\cite{TroutJACS2003} to drive nucleation, and the generation of statistically representative nucleation trajectories that can allow direct and accurate rate predictions has so far been beyond reach. 

\begin{figure}
	\centering
	\includegraphics[width=.5\textwidth]{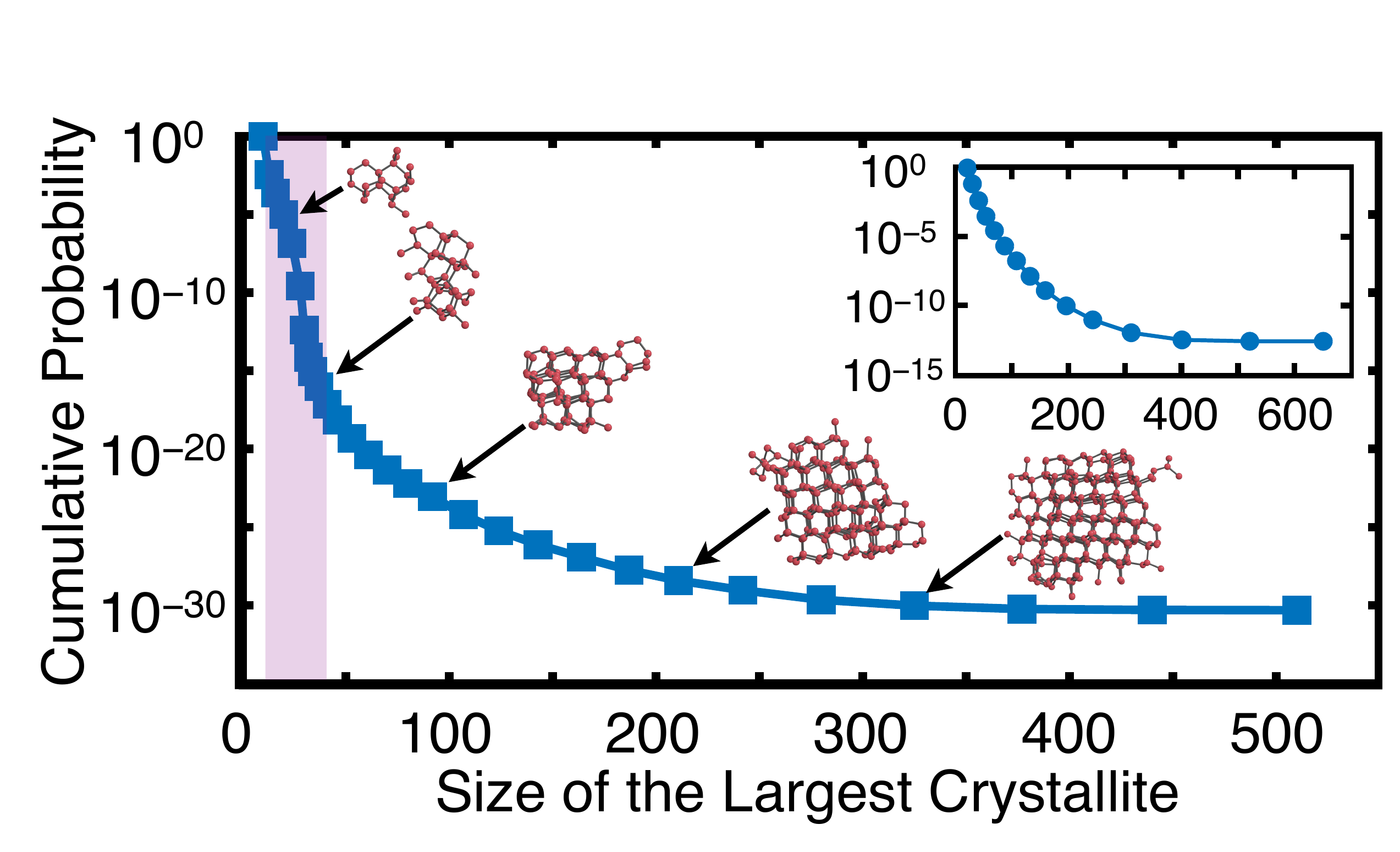}
	\caption{\label{fig:prob}
	\textbf{Cumulative transition probability vs. size of the largest crystalline nucleus in the TIP4P/Ice system at 230~K and 1~bar.} The inflection region is shown in shaded purple. Several representative crystallites are also depicted. The cumulative probability curve for the LJ system simulated at $k_BT/\epsilon=0.82$ and $p\sigma^3/\epsilon=5.68$ is shown in the inset with $\epsilon$ and $\sigma$ the LJ energy and size parameters. No inflection region is observed in the LJ system.
	}
\end{figure}

In this work, we achieve this goal in a system of $4,\!096$ water molecules at 230~K and 1~bar by introducing a novel coarse-graining modification to the path sampling method known as  forward-flux sampling (FFS)~\cite{AllenFrenkel2006}. In the FFS approach, the nucleation process is sampled in stages defined by an order parameter, $\lambda$. In crystallization studies, $\lambda$ is typically chosen as the size of the largest crystalline nucleus in the system~\cite{GalliPCCP2011, GalliNatComm2013, HajiAkbariFilmMolinero2014}. Individual molecules are labeled as solid- or liquid-like based on the Steinhardt order parameters~\cite{SteinhardtPRB1983}, and the neighboring solid-like molecules are connected to form a cluster (For further details, refer to the SI and Fig.~S1.). The cumulative probability of growing a crystallite with $\lambda$ molecules is then computed from the success probabilities at individual stages. If a sufficiently large number of trajectories are sampled at each stage, the nucleation mechanism can be accurately determined by inspecting the ensemble of pseudo-trajectories that connect the liquid and crystalline basins. We use the term 'pseudo-trajectory` as, during FFS, all velocities are randomized at any given milestone.

\begin{figure*}
	\begin{center}
	\includegraphics[width=.8366\textwidth]{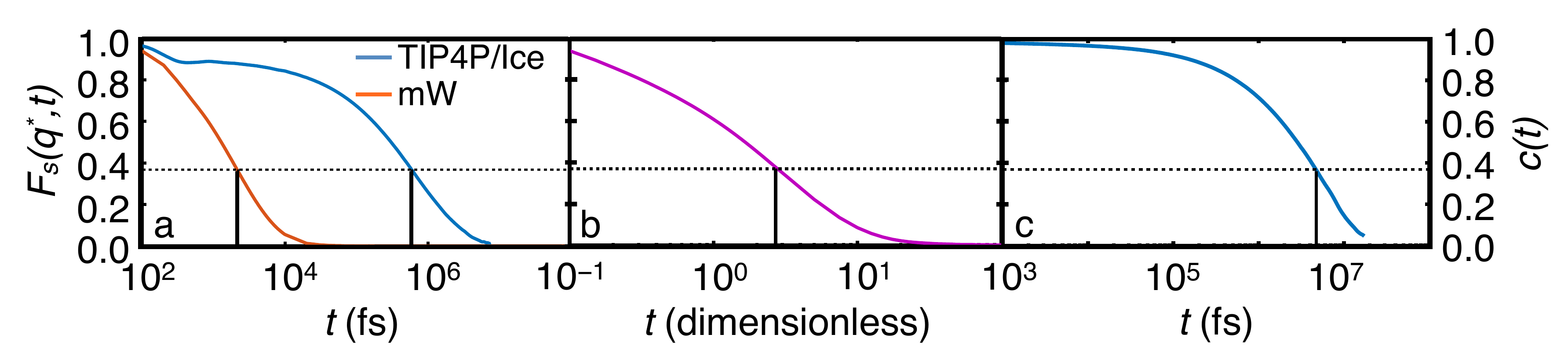}
	\caption{\label{fig:rlxtime} \textbf{Structural relaxation in the supercooled liquid}. Self-intermediate scattering functions computed from MD simulations of (a) the TIP4P/Ice (blue) and the mW (orange) systems at 230~K and 1~bar, and (b) the LJ system at $k_BT/\epsilon=0.82$ and $\rho_N\sigma^3=0.974$. In each case, $q^*$ is in close correspondence with the first peak of $S(q)$, the structure factor, in the corresponding system.  The structural relaxation time, $\tau_r$, is defined as the time at which $F_s(q^*,t)=1/e$. (c) $c(t)$, the hydrogen-bond correlation function, computed in $NpT$ simulations of a system of 216 TIP4P/Ice molecules at 230~K and 1~bar. $\tau_h$ is defined as $c(\tau_h)=1/e$.	}
	\end{center}
\end{figure*}

\begin{figure}
	\begin{center}
		\includegraphics[width=0.4876\textwidth]{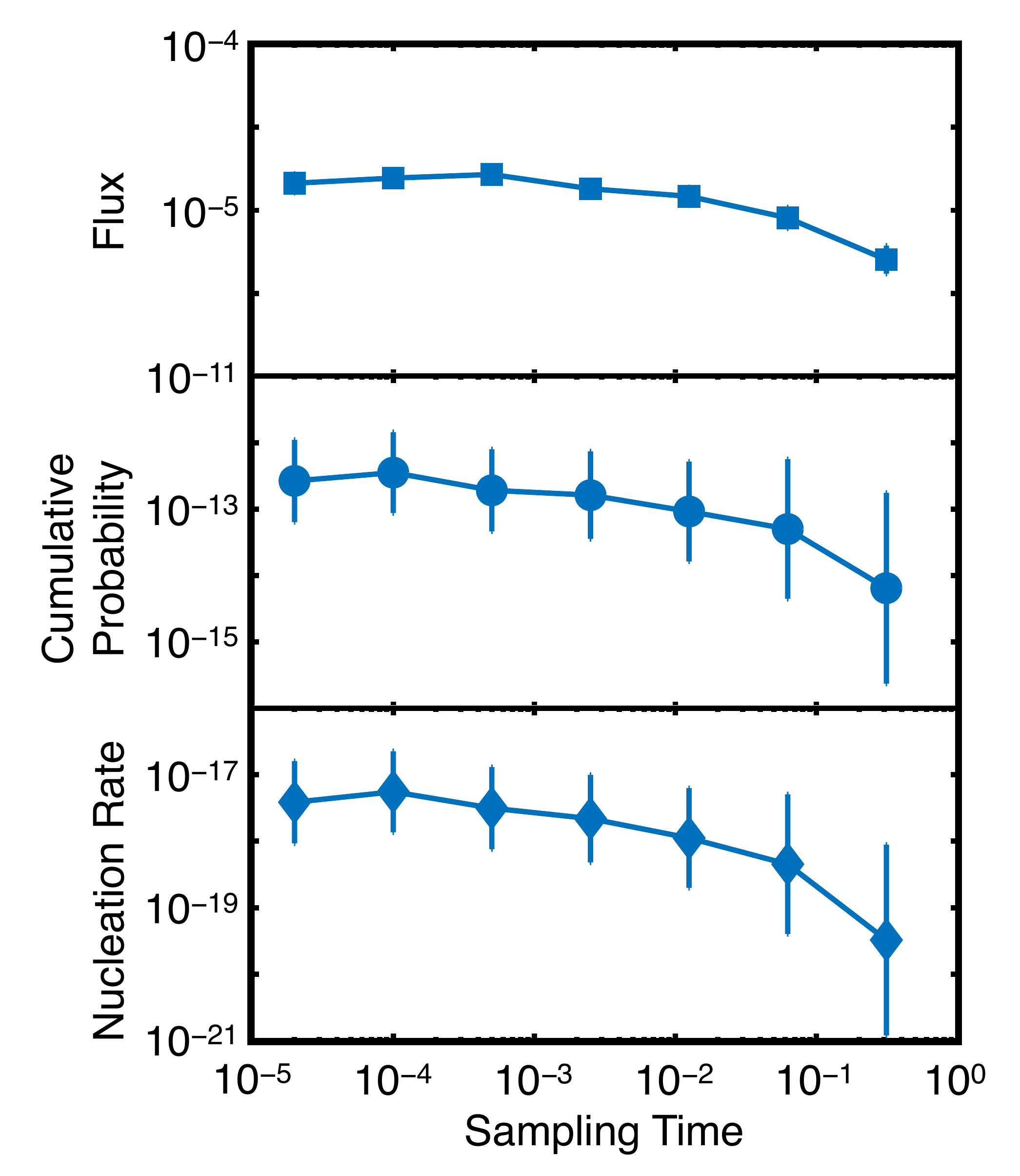}
		\caption{\label{fig:samplingwindow}
		\textbf{Effect of $\tau_s$, the sampling time, on  fluxes, cumulative probabilities and nucleation rates computed from a series of FFS calculations conducted for a system of $4,\!096$ Lennard-Jones atoms at $k_BT/\epsilon=0.82$ and $p\sigma^3/\epsilon=5.68$.} Divergence only occurs when  $\tau_s$ becomes comparable to $\tau_r$. Computed quantities are insensitive to $\tau_s$ for $\tau_s\ll\tau_r$. All quantities are in the LJ dimensionless units.
		}
	\end{center}
\end{figure}

In  conventional FFS, the underlying MD trajectories are monitored as frequently as possible, usually every single MD step. In the TIP4P/Ice system, however, this approach is unsuccessful as the cumulative growth probability never converges (plateaus) and instead plummets unphysically (Fig.~S2a). Because of the five-orders-of-magnitude separation between the structural relaxation time, $\tau_r$ (Fig.~\ref{fig:rlxtime}a), and the sampling time, $\tau_s$, the high-frequency fluctuations in $\lambda(t)$ do not reflect physically relevant structural transformations. We therefore filter such high-frequency fluctuations by computing the order parameter along MD trajectories less frequently. We choose $\tau_s=1$~ps, which is still around three orders of magnitude smaller than the hydrogen bond relaxation time~\cite{LuzarNature1996} (Fig.~\ref{fig:rlxtime}c). By decreasing the separation between $\tau_s$ and $\tau_r$, the FFS calculation converges and the cumulative probability eventually plateaus 
(Fig.~\ref{fig:prob}).  The computed nucleation rate is $\log_{10}R=5.9299\pm0.6538$-- $R$ in m$^{-3}\cdot$s$^{-1}$.  This implies, statistically, one nucleation event per $9\times10^{18}$~s in the $4,\!096$-molecule system considered in this work, that has an average volume of $\approx125$~nm$^3$.  Note the astronomical separation of time scales between structural relaxation ($\tau_r=0.6$~ns) and ice nucleation. This rate is placed in the context of earlier experimental estimates~\cite{TaborekPRB1985,NilssonNature2014} below (see Comparison with Experimental Rate Measurements). We confirm the accuracy of the coarse-grained FFS by observing that the computed crystallization rates in the Lennard-Jones (LJ) system are insensitive to $\tau_s$ if $\tau_s/\tau_r<10^{-1}$ (Fig.~\ref{fig:samplingwindow} and~\ref{fig:rlxtime}b).

\begin{figure*}
	\centering
	\includegraphics[width=0.8960\textwidth]{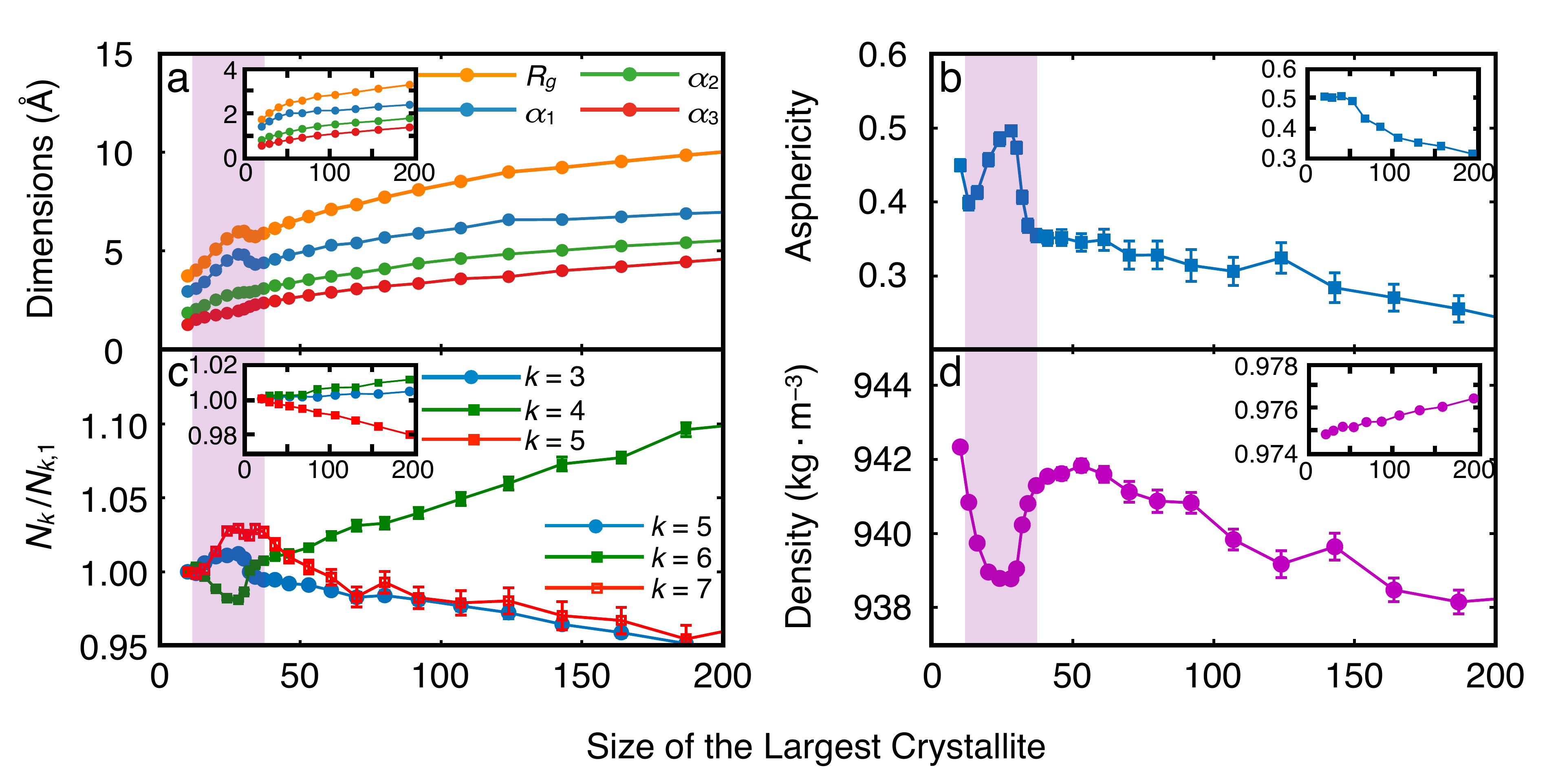}
	\caption{\label{fig:non_mon}
	\textbf{Nonmonotonicities in average mechanical observables for the configurations obtained from the FFS calculation.} The insets correspond to the FFS calculation in the LJ system. (a) Radius of gyration ($R_g$), principal axes ($\alpha_1\ge\alpha_2\ge\alpha_3$) and (b) asphericity of the largest crystallite. (c) Ring statistics and (d) density of the system. $N_k(\lambda)$ corresponds to the average number of $k$-member rings at $\lambda$, with $N_{k,1}=N_k(\lambda_1)$. For water, five-, six- and seven-member rings are enumerated, while for the LJ system, three-, four- and five-member rings are enumerated.   The shaded purple region corresponds to the inflection region. All quantities are in dimensionless units for the LJ system.
	}
\end{figure*}

For most materials, the probability of adding a certain number of molecules to a crystallite of $\lambda$ molecules increases with $\lambda$. This leads to a consistent positive curvature in the cumulative probability curve e.g.~in the crystallization of the LJ system (Fig.~\ref{fig:prob} inset and Fig.~S3a). For water, however, the cumulative probability curve has a pronounced inflection  at  $\lambda\approx30$, where the probability of growing an average crystallite decreases significantly with $\lambda$ before rebounding again at larger $\lambda$'s. The inflection is accompanied by non-monotonicities in several other mechanical observables. For instance, in the inflection region, the average density increases with $\lambda$  (Fig.~\ref{fig:non_mon}d), even though there is an overall decrease in density upon crystallization. We observe similar non-monotonicties in the longest principal axes (Fig.~\ref{fig:non_mon}a) and the asphericity (Fig.~\ref{fig:non_mon}b) of the largest crystallite, as well as the number of five-, six- and seven-member rings in the system (Fig.~\ref{fig:non_mon}c). The non-monotonicity in ring size distribution has also been observed in the freezing of ST2, another molecular model of water~\cite{PalmerNature2014}. In the LJ system, however, all of these quantities evolve monotonically from their averages in the liquid to their averages in the crystal (Fig.~\ref{fig:non_mon} insets and Fig.~S3). In the coarse-grained mW system, this inflection is present, but is very mild, and the non-monotonicities are much weaker (Fig.~S4).

\begin{figure*}
	\centering
	\includegraphics[width=0.8407\textwidth]{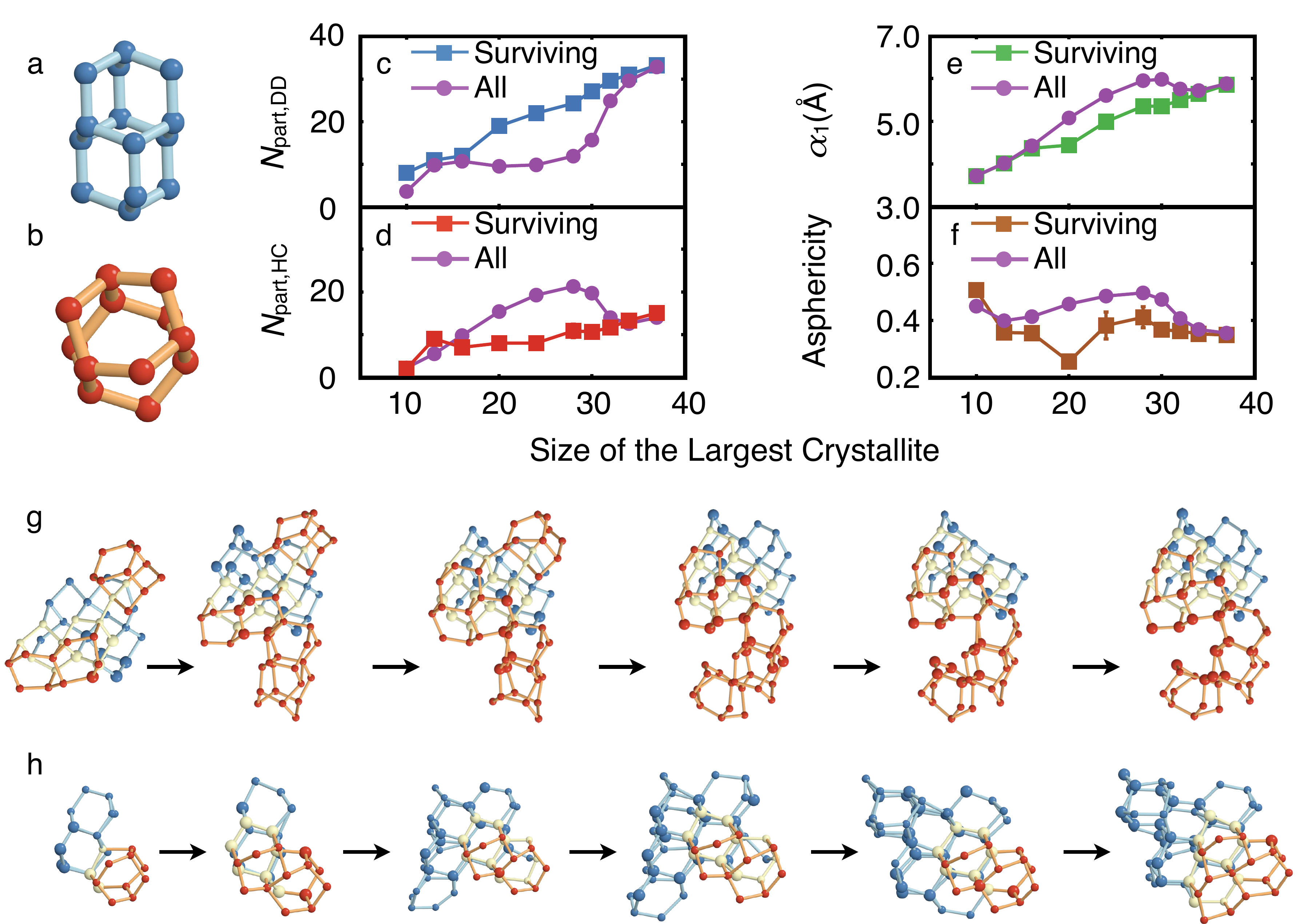}
	\caption{\label{fig:ddc_hc}
	\textbf{Competition between cubic and hexagonal ice in the inflection region.} (a) Double-Diamond, and (b) hexagonal cages. (c-d) Number of water molecules in the largest crystallite that participate in (c) a DDC and (d) an HC. (e) The longest principal axis and (f) asphericity of the largest crystallite. (g) A pseudo-trajectory that does not survive the inflection region. DDC and HC cages  shown in blue and red, respectively. Yellow particles belong to both a DDC and an HC. 
Note  the abundance of HCs. 
(h) A pseudo-trajectory that survives the inflection region. Note the abundance of DDCs. Molecules that are part of the largest crystallite (based on $q_6$) are shown larger than  liquid-like molecules that participate in the topological DDC/HC network that encompasses the largest crystallite. 
	}
\end{figure*}

In order to understand the origin of this inflection, we examine all the configurations in the shaded purple regions of Figs.~\ref{fig:prob} and \ref{fig:non_mon}, and identify those that 'survive` the inflection region by giving rise to a progeny at $\lambda=41$. Visual inspection of these configurations reveals an abundance of double-diamond cages (DDCs) in their largest crystallites. DDCs (Fig.~\ref{fig:ddc_hc}a) are the basic building blocks of cubic ice ($I_c$), and are topologically identical to the carbon backbone of the polycyclic alkane diamantane~\cite{CourtneyJCSPT1972}. The largest crystallites of the 'vanishing` configurations, however, are rich in hexagonal cages (HCs) (Fig.~\ref{fig:ddc_hc}b), the basic building blocks of hexagonal ice ($I_h$). We then use a topological criterion to detect DDCs and HCs (See SI). In this approach, all primitive hexagonal rings in the nearest neighbor network are identified, and DDCs and HCs are detected based on the connectivity of the neighboring hexagonal rings (See SI for further details.). We identify several isolated cages even in the supercooled liquid.  Due to their distorted geometries, however, such cages can only be detected topologically, and not through  conventional order parameters such as $q_3$~\cite{GalliPCCP2011}. 
Similar to the crystallites that are clusters of neighboring molecules with local solid-like environments (see SI), the cages that share molecules can also be clustered together to define interconnected DDC/HC networks. With their constituent cages detected topologically, such networks can contain both solid- and liquid-like molecules. We observe that almost all the molecules of the largest crystallites participate in DDC/HC networks. This is consistent with earlier experimental and computational observations~\cite{SalzmannPNAS2012, MolineroPCCP2011} that the ice that nucleates from supercooled water is a stacking-disordered mixture of both $I_c$ and $I_h$ polymorphs. 

Consistent with our visual observation, a stark difference exists between the DDC makeup of the surviving and vanishing configurations. In the surviving configurations, the water molecules of the largest crystallite are more likely to participate in DDCs than in HCs (Figs.~\ref{fig:ddc_hc}c-d), making the corresponding crystallites more cubic than the average. Such cubic-rich configurations are  scarce at the beginning and only grow in number towards the end of the inflection region. Conversely, the majority of configurations, which are HC-rich, become extinct towards the end of the inflection region. This preference can be explained by comparing the geometric features of the HC-rich and DDC-rich crystallites. While the DDC-rich crystallites are comparatively uniform in shape (Fig.~\ref{fig:ddc_hc}h), the HC-rich crystallites are more aspherical (Fig.~\ref{fig:ddc_hc}g), and therefore less likely to grow and survive the inflection region. This higher asphericity arises from the preferential addition of new HCs to the prismatic faces of the existing HCs, as evident in the abrupt increase in the ratio of prismatic to basal HC-HC connections in the inflection region (Fig.~S5f). This is qualitatively consistent with earlier observations showing that the growth of bulk hexagonal ice is faster along its prismatic plane~\cite{SeoJCP2012}. The preference for cubic ice in the early stages of nucleation has been observed in previous studies of ice formation in different water models~\cite{MolineroPCCP2011, BrukhnoJPhysCondMat2008, QuigleyJCP2008}. However, the molecular origin of this preference had not been identified prior to this work. Indeed, the non-monotonicties in the shape and asphericity of the largest crystallite almost disappear  when only the surviving configurations are considered (Fig.~\ref{fig:ddc_hc}e-f). A similar correlation exists between the DDC makeup of a configuration and its density and ring size distribution (Fig.~S6).

\begin{figure}
	\centering
	\includegraphics[width=0.3721\textwidth]{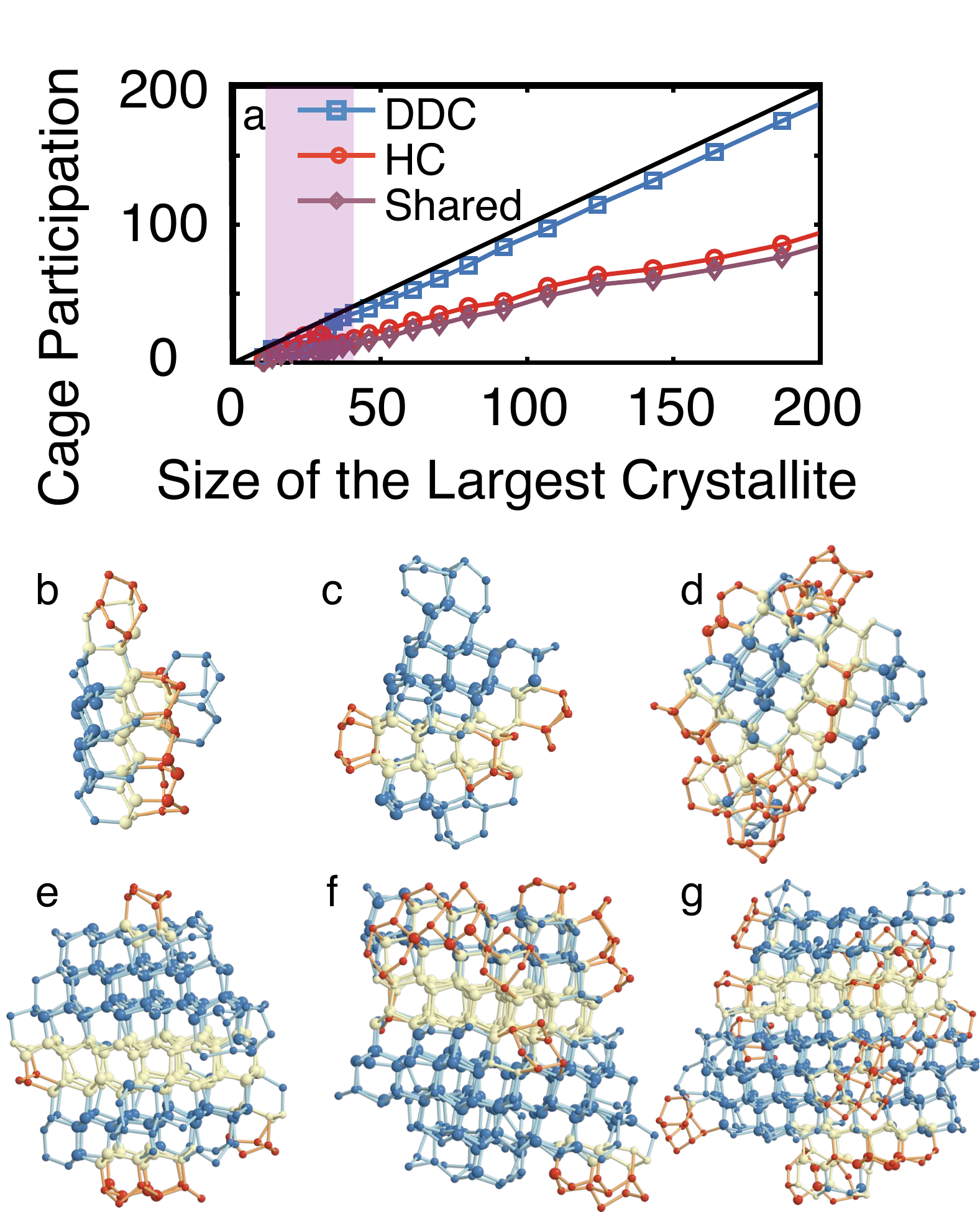}
	\caption{\label{fig:beyond_inflection}
	\textbf{Nucleation beyond the inflection region.} (a) Average cage participation of the molecules in the largest crystallite.  The solid black line has a slope of unity.  The molecules that participate in a DDC (or HC) are included in the corresponding count even if they also participate in a neighboring cage of the other type. The overwhelming majority of molecules are at least part of a DDC, while very few molecules are only a part of an HC. (b-g) Several representative configurations obtained at different milestones after the inflection region. (b-e) are pre-critical, (f) is critical and (g) is post-critical. Molecules that are a part of a DDC, an HC, or both are depicted in dark blue, dark red, and light yellow respectively.  Here, we use the same size convention used in Fig.~\ref{fig:ddc_hc}.
	}
\end{figure}

Fig.~\ref{fig:beyond_inflection} depicts the fate of the cubic-rich crystallites that survive the inflection region. Due to the  thermodynamic stability of $I_h$ relative to $I_c$, one expects the surviving cubic-rich crystallites to eventually transform into $I_h$. We observe no such transformation during the nucleation process, and the crystallites retain their high DDC content (Fig.~\ref{fig:beyond_inflection}a) even  after they are post-critical (Fig.~\ref{fig:beyond_inflection}g). (For a discussion of criticality, wee SI and Fig.~S7b.) This suggests the need for caution in the interpretation of earlier indirect calculations of nucleation rate~\cite{SanzJACS2013} in which the critical nuclei are assumed to be exclusively hexagonal. We also observe no tendency for the hexagonal polymorph to prefer the core of the crystallite. This is in contrast to the traditional picture of nucleation in which the more thermodynamically stable phase concentrates at the core, with a shell of the less stable phase shielding it from the liquid~\cite{FrenkelJCP1996}. Instead, we observe a large number of exposed hexagonal cages at the surface (Figs.~\ref{fig:beyond_inflection}b-g), with attrition tendencies similar to the HCs in the inflection region (e.g.~the HC appendages in Fig.~\ref{fig:beyond_inflection}d and the large prismatic-to-basal ratio in Fig.~S5f). The propensity to grow more cubic stacks even after the inflection region is consistent with the proposed mechanism as the addition of new HCs to a large crystallite is more likely to lead to chain-like appendages at the surface, henceforth making it less stable than an equally-sized crystallite grown via the addition of  DDCs. Indeed, the propensity to form thicker cubic stacks has been observed in the growth and consolidation of post-critical crystallites in the growth-limited freezing of the mW system~\cite{MolineroPCCP2011}.

\section{Comparison with Experimental Rate Measurements}

As mentioned above, experimental measurements of nucleation rate are only practical over a narrow range of thermodynamic conditions. This is because of the fundamental limitation of existing experimental techniques, which are based on probing the temporal evolution of the number of freezing events that are detected in a small population of supercooled micro-droplets~\cite{KoopZPC2004}. Therefore, the nucleation rates that can be measured from the existing experimental techniques can span few orders of magnitude only, confining the range of thermodynamic conditions over which nucleation rates are measurable. For temperatures that are outside this range, the nucleation rate is either so small that none of the micro-droplets would freeze during the timescale of the experiment, or is so large that all droplets would freeze immediately. For droplets as small as a few micrometers in diameter, nucleation rates have been measured for temperatures as low as 234~K, which corresponds to a supercooling of 39~K, 3~K smaller than the supercooling considered in this work. (The melting temperature of the TIP4P/Ice model is 272.2~K~\cite{VegaTIP4PiceJCP2005} vs.~the experimental melting temperature of 273~K. Henceforth, our temperature of 230~K corresponds to a supercooling of 42~K.) Therefore, no direct comparison can be made between our computed nucleation rate and any actual experimental measurement, without extrapolating to lower temperatures. These extrapolations are typically based on classical nucleation theory, and are prone to large uncertainties, leading to large variations in the extrapolated nucleation rates. In particular, such extrapolations fail to take into account the transition to the transport-controlled nucleation at low temperatures, which is responsible for the appearance of a maximum in the nucleation rate with respect to temperature. For real water, the temperature of maximum crystallization rate has been estimated to be $\approx225~$K~\cite{MolineroNature2011}, which is very close to the temperature considered in this work. Such extrapolations yield a wide range of nucleation rates at a supercooling of 42~K, from $10^{18}\text{m}^{-3}\cdot\text{s}^{-1}$ in Ref.~\cite{TaborekPRB1985, DeMottJAtmosSci1990} to $10^{24}\text{m}^{-3}\cdot\text{s}^{-1}$ in Ref.~\cite{HeymsfieldJAtmosSci1993}. Another potential source of error, which can lead to a systematic overestimation of rates at lower temperatures, is the possibility of surface-dominated nucleation in smaller droplets that are typically used for rate measurements at lower temperatures~\cite{TabazadehPNAS2002}. 

\begin{table*}
	\centering
	\caption{\label{tab:expcp}Numerical correlations used for fitting the experimental heat capacity measurements of Refs.\cite{GiauqueJACS1935,AngellJPC1973}. Units are in $\text{cal}\cdot\text{mol}^{-1}\cdot\text{K}^{-1}$. For hexagonal ice, we use a linear fit, while for supercooled water we use a combination of a power law and a linear fit. The actual experimental data for supercooled water are for $T\ge236$~K. We thus use the numerical fit provided below to extrapolate $C_p$ at $231~\text{K}\le{T}\le236~$K.}
	\begin{tabular}{ll}
		\hline\hline
		\sf{Phase} & \sf{Correlation} \\
		\hline
		\sf{Supercooled Water\cite{AngellJPC1973}}~~~~~~~~~~~~~~~~~~~~~~& $C_p(T)=4\times10^{15}T^{-5.824}+0.7131T-202$\\
		\sf{Hexagonal Ice\cite{GiauqueJACS1935}}~~~~~~~~~~~~~~~~~~~~~~ & $C_p(T)=0.032T+0.3252$\\
		\hline
	\end{tabular}
\end{table*}

Thanks to the superior temporal resolution of new experimental techniques, direct measurements of nucleation rates at larger supercoolings will be possible in the near future. One such technique is the femtosecond X-ray laser pulsing that was recently used by Sellberg~\emph{et al.}~\cite{NilssonNature2014} to probe the structural transformation of a population of evaporatively cooled micro-droplets of supercooled water. Although no nucleation  rates are reported in their work, it is possible to obtain an approximate estimate using Fig.~2 of Ref.~\cite{NilssonNature2014}, which depicts the temporal profiles of the temperature and the ice fraction of  evaporatively cooled 12~$\mu$m droplets. The first frozen droplets are detected approximately four milliseconds after they enter the chamber and when they reach a temperature of 229~K. The average freezing time of 4~ms can be used to obtain an upper bound of $J_v\approx2.7631\times10^{17}\text{m}^{-3}\cdot\text{s}^{-1}$ for the homogeneous nucleation rate at the supercooling of 42~K.  This is around eleven orders of magnitude larger than the rate computed in this work. As we will discuss below, however, this discrepancy is reasonable considering the sensitivity of the nucleation rate to different thermodynamic features of the system. According to the classical nucleation theory, the nucleation rate is proportional to $\exp[-\Delta{G}_c/k_BT]$ with $\Delta{G}_c$, the free energy barrier associated with the formation of a critical nucleus, given by:
\begin{eqnarray}
\Delta{G}_c &=& \frac{16\pi\gamma^3}{3\rho_s^2|\Delta\mu|^2}\label{eq:barrier}
\end{eqnarray}
Here $\gamma$ is the solid-liquid surface tension, $\rho_s$ is the number density of the solid, and $\Delta\mu$ is the free energy difference between the crystalline and liquid phases. The exponential dependence of the nucleation rate on $\Delta{G}_c$, and the sensitivity of $\Delta{G}_c$ to $\gamma$ and $\Delta\mu$  implies that only a slight deviation of any of these quantities from the experimental value can shift the nucleation rate by several orders of magnitude. Both these quantities are difficult to measure at large supercoolings, mostly because of the difficulty of stabilizing supercooled water at such low temperatures. 

\subsection{Free Energy Difference} 
If $\Delta{H}_f$, the latent heat of fusion, is not a strong function of temperature, $\Delta\mu$ can be approximated as  $\Delta\mu\approx\Delta{H}_f(T_f-T)/T_f$. This approach, which yields a value of $\Delta\mu\approx0.2215$~$\text{kcal}\cdot\text{mol}^{-1}$ at a supercooling of 42~K, is, however, not very accurate for water due to its heat capacity anomaly. In order to obtain a more accurate estimate, we take the heat capacity measurements for hexagonal ice~\cite{GiauqueJACS1935} and supercooled water~\cite{AngellJPC1973} (Table~\ref{tab:expcp}), and use thermodynamic integration to obtain a more accurate estimate of $\Delta\mu_{\text{exp}}=0.1855$~$\text{kcal}\cdot\text{mol}^{-1}$. Similarly we use MD simulations in the $NpT$ ensemble to compute enthalpies of hexagonal ice and supercooled water at $230$~K$\le{T}\le272$~K and utilize those enthalpies to compute $\Delta\mu$ using thermodynamic integration. We obtain a value of  $\Delta\mu_{\text{TIP4P/Ice}}=0.147$~$\text{kcal}\cdot\text{mol}^{-1}$ for the TIP4P/Ice system, which is around 20 per cent smaller than $\Delta\mu_{\text{exp}}$. This discrepancy alone can lead to an overestimation of the nucleation barrier by as much as 60 per cent if everything else is identical. To be more quantitative, the classical nucleation theory predicts a nucleation barrier of $\Delta{G}_c=\frac12|\Delta\mu|N_c\approx51k_BT$ for the TIP4P/Ice system at $T=230$~K. However, if we use $\Delta\mu_{\text{exp}}$ instead of $\Delta\mu_{\text{TIP4P/Ice}}$, and $\rho_{s,\text{exp}}=0.922~\text{g}\cdot\text{cm}^{-3}$ (Ref.~\cite{MelinderIntJR2010}) instead of $\rho_{s,\text{TIP4P/Ice}}=0.908~\text{g}\cdot\text{cm}^{-3}$ (obtained from $NpT$ MD simulation of $I_h$ at 230~K and 1~bar) in (\ref{eq:barrier}), we obtain a barrier of $\approx31k_BT$, which  corresponds to an increase in the nucleation rate by  8-9 orders of magnitude. This is very close to the discrepancy between our calculation and the experimental estimates of rate.

\subsection{Surface Tension}
At temperatures below $T_f$, the supercooled water that is in contact with ice is not stable and will immediately freeze. This makes experimental measurements of $\gamma$ in the supercooled regime extremely challenging. Therefore, $\gamma$ is typically estimated indirectly from the nucleation data assuming the validity of the classical nucleation theory. Consequently, there is a large variation in the reported estimates of $\gamma$ for supercooled water that span between 25 and 35~mJ$\cdot\text{m}^{-2}$ (Ref.~\cite{JamesJCP2002}). Similarly, it is very challenging to compute $\gamma$ directly from molecular simulations at $T<T_f$, and all the existing estimates are obtained from nucleation calculations~\cite{GalliPCCP2011, SanzJACS2013, LimmerJCP2012}. The existing direct calculations have all been performed at coexistence conditions~\cite{HandelPRL2008,MolineroJACS2014}. The computed numbers  cover even a wider range from $20.4$~$\text{mJ}\cdot\text{m}^{-2}$ in Ref.~\cite{SanzJACS2013} to $35$~$\text{mJ}\cdot\text{m}^{-2}$ in Ref.~\cite{LimmerJCP2012}. This large variability underscores the sensitivity of the computed value to the particulars of the water model, and to the thermodynamic conditions at which the calculation has been made.  In light of the mechanism that is proposed for freezing in this work, the problem of determining $\gamma$ is further compounded by the stacking disorder nature of the critical nucleus. Considering the cubic dependence of the nucleation barrier on $\gamma$, even the slightest deviation from the experimental value can shift the nucleation rate by several orders of magnitude. For instance, a seven percent deviation can change the nucleation barrier by as much as 22 percent which can shift the nucleation rate by several orders of magnitude.

Overall, the existing classical models of water inevitably predict certain thermodynamic properties of water to be at variance with experiments by a significant margin, and a model that predicts all thermodynamic properties accurately is yet to be developed~\cite{VegaPCCP2011}. Therefore,  the agreement between the orders of magnitude of the computational estimate of the nucleation rate in a classical model of water, like TIP4P/Ice, and the corresponding experimental value is difficult to achieve with the existing models due to strong sensitivity of the nucleation rate to the particular thermodynamic features of the employed water model (e.g.~the free-energies of the liquid and the solid, and the liquid-solid surface tension). 

\subsection{Earlier Computational Studies of Nucleation Rate}  It is inherently problematic to compare  computational estimates of  nucleation rate obtained for different force-fields using different methodologies. This is not only because of the large uncertainties associated with the utilized methods (e.g.~the validity of  classical nucleation theory in the seeding technique), but also due to the empirical and approximate nature of the utilized force-fields, which, as shown in the previous section, can shift the computed rates by several orders of magnitude. This difficulty becomes apparent upon observing the spread of the reported computational estimates of the homogeneous ice nucleation rate in the literature.  Li~\emph{et al}~\cite{GalliPCCP2011, GalliNatComm2013} and Haji-Akbari~\emph{et al}~\cite{HajiAkbariFilmMolinero2014} used forward-flux sampling to compute the nucleation rate in the mW system over a wide range of temperatures. Their computed rates are lower than the corresponding experimental values, but are yet a few orders of magnitude higher than the rate computed in this work. This can not only be attributed to the inherently faster dynamics of the mW system, but also to the higher $|\Delta\mu|$ of the mW model at deep supercoolings, as depicted in Fig.~2a of Ref.~\cite{VegaJCP2014}. Recently, Sanz~\emph{et al.}~\cite{SanzJACS2013, VegaJCP2014} used the seeding technique to compute the nucleation rates for several water models. This interesting approach assumes the validity of the classical nucleation theory and the precise crystallographic nature of the critical nucleus (e.g. hexagonal ice). The estimated uncertainties associated with these-- and other-- assumptions are very large (e.g. error bars in Fig.~7 of Ref.~\cite{SanzJACS2013}). But nevertheless, a comparison at the same reduced conditions suggests that the present rates are lower than the those estimated by Sanz~\emph{et al}. In particular, the rates reported at Refs.~\cite{SanzJACS2013, VegaJCP2014} are \ very sensitive to the \emph{a priori} definition of what constitutes a nucleus, as the size of the critical nucleus is directly used for estimating the nucleation barriers and the nucleation rates. Our approach, however, does not rely on determining the size of the critical nucleus as a prerequisite for computing the nucleation rate.

\section{Conclusions}

In this work, we establish the feasibility of computing the rate of homogeneous ice nucleation for  realistic molecular models of water.
This is significant considering the difficulties associated with measuring  nucleation rates in experiments. However, the computed rates for the TIP4P/Ice model are several orders of magnitude smaller than the experimental estimates at comparable conditions. This discrepancy is attributed to the smaller thermodynamic driving force for the freezing of the TIP4P/Ice system relative to experiment. Nevertheless, the ability to directly compute rates for a molecular model makes it possible, in principle, to parameterize molecular force-fields with an eye towards accurate prediction of nucleation rates. In addition, this paves the way for studying the kinetics and mechanism of ice nucleation across a wide range of  environments, such as the atmospherically relevant films, droplets and aerosols. Finally, the coarse-grained FFS utilized in this work can  prove useful in studying disorder-order transitions in other slowly-relaxing systems, such as water/gas mixtures, ionic liquids, and macromolecular and biomolecular systems. In addition to being able to compute nucleation rates, we obtain valuable mechanistic information that is not attainable in experiments. In particular, we provide a molecular explanation for the initial formation of cubic-rich ice in homogeneous nucleation of supercooled water. 

\section{Methods}
Individual MD simulations are performed using LAMMPS~\cite{PimptonLAMMPS1995}. The size of the largest crystalline nucleus is chosen as the order parameter. The largest crystallites are detected using the Steinhardt $q_6$ order parameter~\cite{SteinhardtPRB1983} and the chain exclusion algorithm of Reinhardt~\emph{et al.}~\cite{VegaJCP2012}. Technical specifications of the MD simulations and the order parameter can be found in the SI. Rings are detected using the King criteria~\cite{KingNature1967}, while DDCs and HCs are identified using a novel topological approach, with the detection algorithms thoroughly mentioned in the SI. 

Rate calculations are performed applying a novel coarse-graining to the forward flux sampling algorithm~\cite{AllenFrenkel2006}. The FFS technique is based on sampling the nucleation process in stages, by staging milestones between the liquid and crystalline basins. (See SI for further explanation.) The essence of FFS is thus to identify first passage events between the absorbing milestones of each iteration. In principle, this should be done by monitoring all the time-continuous trajectories originating at any given milestone and by determining the exact times at which they cross any of the two absorbing milestones. In reality, however, these time-continuous trajectories are approximated by solving the discretized versions of the equations of motion. As a result, the order parameter can only be computed as frequently as every single MD step, and any crossings that might occur at intermediate times will be inevitably ignored. Historically, this has been the approach taken in all reported applications of the FFS algorithm, with some authors using larger sampling times (up to a few MD steps) only out of convenience~\cite{LiNatMater2009}.  In the context of crystallization, however, it is reasonable to argue that  fluctuations in the order parameter are only meaningful if they occur at time scales that are not significantly smaller than the structural relaxation time, $\tau_r$, or the hydrogen-bond relaxation time, $\tau_h$. One can therefore coarse-grain the FFS algorithm by using a larger sampling time, and ignoring any high-frequency oscillations in the order parameter that occur at intermediate times. In order to test the validity of this argument, we carry out a series of FFS calculations of the rate of homogeneous crystal nucleation in the LJ system, with sampling times spanning over four orders of magnitude.  We confirm that the cumulative probabilities and nucleation rates are virtually insensitive to the selection of $\tau_s$ unless $\tau_s/\tau_r>10^{-1}$ (Fig.~\ref{fig:samplingwindow}). This approach, however, leads to considerable errors when $\tau_r\sim\tau_s$ as the system starts losing some of its memory between successive samplings of the trajectory. 
No loss of physically relevant information occurs when $\tau_s\ll\tau_r$ as the fluctuations of the order parameter at times smaller than $\tau_s$ are not representative of physically relevant structural transformations. However, these high-frequency fluctuations can become an issue at extremely small sampling times, as in the TIP4P/Ice system. Choosing a large sampling time is, thus, conceptually similar to applying a low-pass filter to the order parameter time series.  In the TIP4P/Ice system, we choose a sampling time of 1~ps, which is 2--3 orders of magnitude smaller than both $\tau_r=0.6$~ns (Fig.~\ref{fig:rlxtime}a) and $\tau_h=4.0$~ns (Fig.~\ref{fig:rlxtime}c). By doing this, we manage to turn an otherwise diverging unsuccessful FFS calculation (Fig.~S2a) into a converging successful calculation presented in  Fig.~\ref{fig:prob}. Further technical details about the method, as well the computational cost of the calculations are included in the SI.

\acknowledgments
P.G.D. gratefully acknowledges the support of the National Science Foundation (Grant No. CHE-1213343) and of the Carbon Mitigation Initiative at Princeton University (CMI). These calculations were  performed on the Terascale Infrastructure for Groundbreaking Research in Engineering and Science (TIGRESS) at Princeton University. This work also used the Extreme Science and Engineering Discovery Environment (XSEDE), which is supported by National Science Foundation grant number ACI-1053575.  The Blue Gene/Q supercomputer resources for this work were provided 
by the Center for Computational Innovations (CCI) at Rensselaer Polytechnic Institute. The authors gratefully acknowledge I. Cosden for his assistance in securing computational resources for this calculation.

\nocite{LJProcRSoc1924, VegaTIP4P2005, LeknerPhysicaB1997, HaywardJCP1997, AlderMDJCP1959, SwopeJCP1982, NoseMolPhys1984, HooverPhysRevA1985, ParrinelloJAppPhys1981, Hockney1989, RyckaertJCompPhys1977, QuigleyJCP2008, TroutJACS2003, SciortinoJCP1989, ChandraJPCB2003, DelagoJCP2008, EscobedoJCP2007}

\bibliographystyle{pnas}
\bibliography{References}   

\eject\phantom{}
\newpage

\setcounter{figure}{0}
\setcounter{table}{0}
\renewcommand{\thefigure}{S\arabic{figure}}
\renewcommand{\thetable}{S\arabic{table}}

\appendix 
\section{SUPPLEMENTARY INFORMATION}

\subsection{System Preparation and Molecular Dynamics Simulations}

All simulations are carried out in cubic boxes with periodic boundaries. In this work, we study crystallization in three different systems: (i) the TIP4P/Ice~\cite{VegaTIP4PiceJCP2005} system as one of the best non-polarizable molecular representations of water, (ii) the mW system~\cite{MolineroJPCB2009} as one of the most popular and widely used coarse-grained representations of water, and (iii) the Lennard-Jones system~\cite{LJProcRSoc1924} as a prototypical simple liquid. In every system, we choose the thermodynamic state points so that the critical nucleus contains between 300 and 500 molecules. Consequently, a system of $4,\!096$ molecules is sufficient for studying crystallization in all these systems. We choose TIP4P/Ice over the closely-related-- and widely-used-- TIP4P/2005~\cite{VegaTIP4P2005} model because of its more realistic melting temperature, giving rise to slightly higher diffusivities at identical supercoolings. Furthermore, the TIP4P/Ice model provides the most quantitatively accurate prediction of the coexistence lines between different ice polymorphs. We prepare our starting configurations from a dilute simple cubic lattice, and gradually compress it to the target temperature and pressure using $NpT$ MD simulations. We then equilibrate the system at the target thermodynamic conditions for a minimum of $10^3\tau_r$ with $\tau_r$ the structural relaxation time for each system. The starting configurations of the cubic and hexagonal ice are prepared using the unit cells proposed by Lekner~\cite{LeknerPhysicaB1997} and Hayward and Reimers~\cite{HaywardJCP1997}, respectively.

All Molecular Dynamics (MD)~\cite{AlderMDJCP1959} simulations are performed in the isothermal-isobaric ($NpT$) ensemble using LAMMPS~\cite{PimptonLAMMPS1995}. We integrate Newton's equations of motion using the velocity Verlet algorithm~\cite{SwopeJCP1982}. Temperature and pressure are controlled using a Nos\'{e}-Hoover thermostat~\cite{NoseMolPhys1984,HooverPhysRevA1985} and a Parrinello-Rahman barostat~\cite{ParrinelloJAppPhys1981}. In the TIP4P/Ice system, long-range electrostatic interactions are computed using the particle-particle particle-mesh algorithm~\cite{Hockney1989} with a short-range cutoff of $8.5$~\AA. Also, the rigidity of water molecules is enforced using the SHAKE algorithm~\cite{RyckaertJCompPhys1977}. Table~\ref{tab:spec} gives the technical specifications of the MD trajectories  for the three systems studied in this work.

\subsection{Order Parameter\label{section:order}}
We quantify the progress of crystallization using the two-step process explained in detail in our earlier publication~\cite{HajiAkbariFilmMolinero2014}. First, every molecule in the system is labelled as solid-like or liquid-like based on its local environment, with the selection criterion being different from system to system (see below). In this work, we use spherical harmonics to distinguish liquid- and solid-like molecules~\cite{SteinhardtPRB1983}. Then, a graph is constructed by connecting all solid-like neighbors that are within the first nearest-neighbor shell of one another. The number of molecules in the largest connected sub-domain of this graph is used as the order parameter in our FFS calculations. Below we provide the particular criteria used for identifying solid-like molecules in each system.

\noindent
\subsubsection{TIP4P/Ice and mW}
\begin{figure*}[hcp]
	\centering
\begin{sansmath}
	\includegraphics[width=0.8259\textwidth]{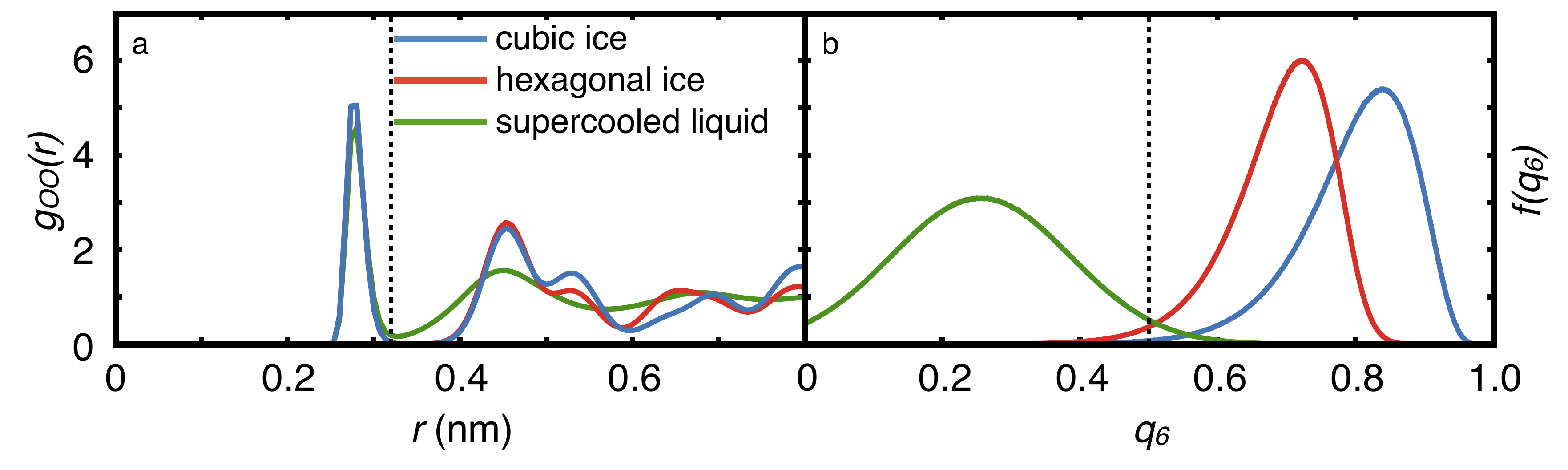}
	\caption{\label{fig:q6} \textbf{Calibration of the order parameter:} (a) Oxygen-oxygen radial distribution function and (b) the distribution of the $q_6$ order parameter  for the cubic and hexagonal polymorphs of ice, and for the supercooled liquid, computed from a $20$-ns $NpT$ MD simulation of the TIP4P/Ice system at 230~K and 1~bar. The distance and $q_6$ cutoffs, $r_c=3.2$~\AA~and $q_{6,c}=0.5$ are both marked with dark dashed lines. }
	\end{sansmath}
\end{figure*}

A distance cutoff of $r_c=3.2$~\AA~is used for defining the first nearest neighbor shell. In the TIP4P/Ice system, $r_c$ corresponds to the location of the first minimum of $g_{OO}(r)$, the oxygen-oxygen radial distribution function, of the disordered, cubic and hexagonal phases (Fig.~\ref{fig:q6}a).
Therefore, the distance between two molecules is defined as the distance between their constituent oxygen atoms. In the mW system, $r_c$ corresponds to the first minimum of $g(r)$~\cite{GalliPCCP2011}. The $q_{6}(i)$ order parameter is then computed for molecule $i$ as:
\begin{eqnarray}
q_6(i) &=& \frac1{N_b(i)}\sum_{j=1}^{N_b(i)}\frac{\textbf{q}_6(i)\cdot\textbf{q}^*_6(j)}{|\textbf{q}_6(i)||\textbf{q}_6(j)|}\label{eq:q6}
\end{eqnarray}
with $\textbf{q}_6(i)\equiv\left(q_{6,-6}(i),q_{6,-5}(i),\cdots,q_{6,6}(i)\right)$  given by:
\begin{eqnarray}
q_{6m}(i) &=& \frac1{N_b(i)}\sum_{j=1}^{N_b(i)} Y_{6m}(\theta_{ij},\phi_{ij}),~-6\le{m}\le6\label{eq:q6m}
\end{eqnarray}
Here $N_b(i)$ is the number of molecules that are within the nearest neighbor shell of the $i$th molecule, $\theta_{ij}$ and $\phi_{ij}$ are the spherical angles associated with the displacement vector between $i$th and $j$th molecules, and $Y_{lm}(\theta,\phi)$ is the $lm$-th spherical harmonic. As depicted in Fig.~\ref{fig:q6}b, a cutoff value of $q_{6,c}=0.5$ is suitable for distinguishing solid- and liquid-like molecules. In order to eliminate chains of locally tetrahedral water molecules, we apply the recently developed algorithm of Reinhardt~\emph{et al}~\cite{VegaJCP2012}. Such chains are abundant in supercooled water, and the failure in removing them can lead to the detection of  non-compact crystallites. As shown in Ref.~\cite{VegaJCP2012}, the application of this chain exclusion step is pivotal in driving crystallization in molecular models of water even when a biasing potential is utilized.

\subsubsection{Hydrogen Bonding and Distance Cutoffs}
Using a distance criterion for identifying solid- and liquid-like molecules in supercooled water is a common practice in the literature~\cite{SanzJACS2013, QuigleyJCP2008}. Alternatively, one can use the hydrogen bond network~\cite{TroutJACS2003}. In deeply supercooled water, however, these two approaches are almost equivalent as the molecules that are within the first neighbor shell of one another are almost always hydrogen-bonded~\cite{SciortinoJCP1989}. In order to quantify this, we identify all the hydrogen bonds in the system. A distance-based bond is a hydrogen bond if the distance between the potential donor oxygen and the acceptor hydrogen is less than 2.42~\AA~and the angle between the O$-$H and the O$\cdots$O vectors is less than 30 degrees~\cite{ChandraJPCB2003}. We observe that more than 98 per cent of distance-based bonds in the entire system are hydrogen bonds. This corresponds to a false positive of less than one bond per every 20 molecules. For the distance-based bonds emanating from the molecules that are part of the largest crystallite, this fraction is $>99.9$ per cent. We are therefore confident that our distance-based criterion that is used not only in computing the order parameter, but also in identifying rings, DDCs and HCs, is robust, and our main conclusions will be unchanged if the hydrogen bond network is used instead.

\subsubsection{Lennard-Jones} We use a distance cutoff of $r_c=1.40\sigma$ as suggested in Ref.~\cite{DelagoJCP2008}. We first compute $\textbf{q}_6(i)$ for each molecule $i$ using Eq.~(\ref{eq:q6m}).
We then carry out an additional level of neighbor averaging as proposed in Ref.~\cite{DelagoJCP2008}:
\begin{eqnarray}
\overline{\textbf{q}}_6 (i) = \frac{1}{N_b(i)}\sum_{j=0}^{N_b(i)}\textbf{q}_6(j)
\end{eqnarray}
with $j=0$ corresponding to the $i$th particle itself. The scalar $q_6$ order parameter is defined as:
\begin{eqnarray}
\overline{q}_6(i) &=& \sqrt{\frac{4\pi}{13}\sum_{m=-6}^6 |\overline{q}_{6m}|^2}
\end{eqnarray}
 Based on Fig.~1(b) of Ref.~\cite{DelagoJCP2008} we use a cutoff of $\overline{q}_{6,c}=0.3$ to distinguish solid- and liquid-like LJ atoms.

\subsection{Forward-flux Sampling\label{section:ffs}}

The forward-flux sampling (FFS) algorithm~\cite{AllenFrenkel2006} is based on partitioning the configuration space into non-overlapping regions separated by milestones that are isosurfaces of the order parameter~\cite{HajiAkbariFilmMolinero2014}. The first milestone, denoted by $\lambda_{\text{basin}}$,  is chosen in the middle of the liquid basin and is therefore frequently crossed by a typical MD trajectory of the liquid. The other milestones $\lambda_{\text{basin}}=\lambda_0<\lambda_1<\lambda_2<\cdots$ are chosen so that $\lambda_k$ is accessible to the trajectories initiated at $\lambda_{k-1}$ with a sufficiently large probability. We have outlined the criteria used for placing $\lambda_{\text{basin}}$ and $\lambda_i$'s in our earlier publication~\cite{HajiAkbariFilmMolinero2014}.  The nucleation rate is then computed as $R=\Phi_0\prod_{k=1}^NP(\lambda_{k+1}|\lambda_k)$. Here, $\Phi_0$ is number of trajectories per unit volume per unit time that cross $\lambda_1$ after crossing $\lambda_{\text{basin}}$ and is calculated by analysing a long MD trajectory in the liquid basin. The configurations that correspond to such crossings are stored for future iterations. The next step is to compute $\{P(\lambda_{k+1}|\lambda_k)\}_{k=1}^N$, the transition probabilities. In order to compute $P(\lambda_{k+1}|\lambda_k)$, a large number of MD trajectories are initiated from the configurations gathered at $\lambda_k$ by randomly picking a configuration one at a time, and randomising its momenta according to the Boltzmann distribution. $P(\lambda_{k+1}|\lambda_k)$ is the fraction of those trajectories that cross $\lambda_{k+1}$ prior to crossing $\lambda_{\text{basin}}$ in the opposite direction, and the configurations that correspond to such crossings are stored for future iterations. This procedure is repeated until a value of $\lambda_N$ is reached for which $P(\lambda_{N+1}|\lambda_N)=1$ for \emph{every} $\lambda_{N+1}>\lambda_N$.  The error bars in the flux and the transition probabilities are estimated using the procedure outlined in Ref.~\cite{AllenFrenkel2006}. The rationale behind the coarse-graining of FFS is explained in the Methods section in the main text.

\subsubsection{Technical Specifications of FFS in the TIP4P/Ice System}
We perform our FFS calculations using an in-house C++ program  discussed in detail elsewhere~\cite{HajiAkbariFilmMolinero2014}. This computer program links against the LAMMPS static library and uses it as its internal MD engine. The basin simulations are carried out for $694$~ns, and the first two milestones are placed at $\lambda_{\text{basin}}=5$ and $\lambda_1=10$. We start the second stage of the algorithm after gathering $1,\!685$ configurations at $\lambda_1=10$. During the FFS iterations in the inflection region, we demand a minimum of $\approx1,\!000$ crossings at every milestone. After the inflection region, however, we terminate each iteration after  $\approx300-350$ crossings. Fig.~\ref{fig:trajTime}a depicts the average success and failure times for the trajectories that are initiated at each milestone. These are the average times that it takes for a trajectory to cross the next milestone and to return to the liquid basin, respectively.  Note that the average success and failure times are very large at the latest stages of the FFS algorithm, of the order of tens to hundreds of nanoseconds. This is consistent with earlier observations by Reinhardt~\emph{et al}~\cite{VegaJCP2012} that in molecular models of water, the critical nucleus (as obtained from umbrella sampling simulations in their work) might not grow or shrink even after tens of nanoseconds of regular MD.

\subsubsection{Critical Nucleus Size} 
The systematic way of determining the size of the critical nucleus is to perform committor analysis~\cite{EscobedoJCP2007}. The committor probability of any given configuration is the probability that a random MD trajectory initiated from that configuration crystallizes before returning to the liquid basin. For the critical nucleus, the committor probability is exactly one half as a critical nucleus is equally likely to melt or to grow. Therefore, the critical nucleus size is the average size of the largest crystallites in the ensemble of all critical configurations. The problem of determining committor probabilities for a family of 'suspected` critical nuclei is, however, computationally expensive and can take up to several months even on a large supercomputer. Since the notion of criticality only plays a descriptive role in our discussion, we  use an alternative approach by assuming that the order parameter utilized in our FFS calculations is a good reaction coordinate. Therefore, the committor probability can be approximated as $p_C(\lambda_k)=\prod_{j=k}^{N-1}P(\lambda_{j+1}|\lambda_j)$ with $\lambda_N$ being the final milestone of the FFS calculation. Fig.~\ref{fig:trajTime}b gives $p_C(\lambda)$ vs.~$\lambda$ for the FFS calculation of the nucleation rate in the TIP4P/Ice system. The critical nucleus has around $320\pm20$ water molecules.

\subsubsection{Computational Cost}
As can be seen in Fig.~\ref{fig:trajTime}a, the average failure and success times are very large at the final iterations of our FFS calculations. As a result, we carried out a total of $608~\mu$s of MD trajectories. This amounts to a total of $21,\!452,\!433$ CPU-hours on the Texas-based Stampede supercomputer. Due to the embarrassingly parallel nature of the FFS algorithm, we were able to distribute this very costly calculation across the following supercomputers: the Princeton-based Della and Tiger supercomputers, the TACC-based\footnote{TACC: Texas Advanced Computing Center} Stampede supercomputer, the SDSC-based\footnote{SDSC: San Diego Supercomputer Center} Gordon supercomputer, and the RPI-based\footnote{Rensselaer Polytechnic Institute} Blue Gene/Q supercomputer.

\subsection{Shape of the Largest Crystallite}
We use $\{\textbf{r}_i\}_{i=1}^N$, the positions of the oxygen atoms in the largest crystallite to compute its gyration tensor $\mathcal{G}:=(1/N)\sum_{i=1}^N[\textbf{r}_i-\textbf{r}_{CM}][\textbf{r}_i-\textbf{r}_{CM}]^T$ with $\textbf{r}_{CM}=(1/N)\sum_{i=1}^N\textbf{r}_i$. $\mathcal{G}$ has three real eigenvalues $\alpha_1^2\ge\alpha_2^2\ge\alpha_3^2$. The asphericity, $\kappa$ is computed as:
\begin{eqnarray}
\kappa^2 &=& \frac{3}{2}\frac{\alpha_1^4+\alpha_2^4+\alpha_3^4}{(\alpha_1^2+\alpha_2^2+\alpha_3^2)^2}-\frac12
\end{eqnarray}

\subsection{Ring Statistics\label{section:ring}}
The first step in identifying rings is to define a nearest neighbor network. For this purpose, we use the same distance criterion utilized in the definition of the $q_6$ order parameter. As explained earlier, over 98 per cent of nearest neighbor pairs are indeed hydrogen bonded. We then use the criteria proposed by King~\cite{KingNature1967} to identify all the primitive rings in the system, and get rid of rings that share more than three consecutive water molecules. In the TIP4P/Ice and mW systems, we identify rings of up to eight molecules, while in the Lennard-Jones system, we confine ourselves to rings of six atoms or less. The hexagonal rings detected in the TIP4P/Ice and mW systems are then used for detecting double-diamond and hexagonal cages. 

\subsubsection{Double-diamond Cages} 
We loop through all the hexagonal rings in the system and use the topological features of a DDC (Fig.~\ref{fig:dd}a) to detect double-diamond cages as follows. A hexagonal ring, $R=(m_1,m_2,\cdots,m_6)$-- with $m_k$ denoting the $k$th molecule that is part of the ring-- is the equatorial ring of a DDC if the following conditions are satisfied:
\begin{enumerate}
	\item For every $1\le{k}\le6$, a minimum of three other hexagonal rings pass through $m_k$.
	\item For every triplet $T_k=(m_k,m_{k\oplus1},m_{k\oplus2})$, there is at least one hexagonal ring other than $R_0$ that passes through $m_k,m_{k\oplus1}$ and $m_{k\oplus2}$. Here $a\oplus{b}=(a+b)\mod6$.
	\item If $\{R_{k,j}\}_{j=1}^{n_k}$ are the hexagonal rings other than $R_0$ that pass through $T_k$, there must exist $j_1,j_2,\cdots,j_6$ so that $R_{1,j_1}\cap{R}_{3,j_3}\cap{R}_{5,j_5}\neq\emptyset$ and $R_{2,j_2}\cap{R}_{4,j_4}\cap{R}_{6,j_6}\neq\emptyset$. Also $R_{1,j_1}\cap{R}_{3,j_3}$, $R_{3,j_3}\cap{R}_{5,j_5}$, etc must all have three molecules. 
\end{enumerate}

\noindent\subsubsection{Hexagonal Cages} The two hexagonal rings $R_1=(l_1,l_2,\cdots,l_6)$ and $R_2=(m_1,m_2,\cdots,m_6)$ can be the basal planes of a hexagonal cage (Fig.~\ref{fig:dd}b) if the following conditions are satisfied:
\begin{enumerate}
	\item $R_1\cap{R}_2=\emptyset$.
	\item There exists  $1\le{k}\le6$ so that $m_k$ is a neighbor of $l_1$ or $l_2$, as defined based on the distance criterion.
	\item If $m_k$ is a neighbor of $l_1$, $m_{k\oplus2}$ and $m_{k\oplus4}$ must be  neighbors of $l_3$ and $l_5$ (or $l_5$ and $l_3$), respectively. Adjusting the algorithm to the case of $m_k$ being a neighbor of $l_2$ is straightforward.
\end{enumerate}

\noindent\subsubsection{DDC/HC Networks} The cages that share a minimum of one water molecule are clustered together to form a network of interconnected DDC/HC cages. Due to the topological criterion used in the identification of their constituent cages, such networks can have water molecules that might be detected as 'liquid-like` when the $q_6$ order parameter is used. This explains the presence of liquid-like molecules in the DDC/HC networks shown in Figs.~3 and 4. 

\noindent\subsubsection{Basal and Prismatic Connection}
We determine the type of connection between two neighboring hexagonal cages based on the number of water molecules that they share. As can be seen in  Fig.~\ref{fig:dd}d, two hexagonal cages that are connected through their basal plane have \emph{six} water molecules in common, while two cages that are connected through their prismatic planes (Fig.~\ref{fig:dd}e) have \emph{four} molecules in common.

\subsection{Thermodynamic Integration}
For computing the free energy difference between supercooled water and hexagonal ice, MD simulations of both systems are performed in the $NpT$ ensemble. Each such simulation is comprised of an initial equilibration period of one nanosecond followed by a five-nanosecond production run. The time averages of enthalpies are computed for $230~\text{K}\le T\le272~\text{K}$ during the production runs and are then used for computing the free energy difference using the following equation:
\begin{eqnarray}
\frac{\mu_{\text{hex}}(T)-\mu_{\text{liq}}(T)}{T} &=& \int_{T}^{T_f}\frac{h_{\text{hex}}(\overline{T})-h_{\text{liq}}(\overline{T})}{\overline{T}^2}d\overline{T}
\end{eqnarray}
Here, $h_X$ corresponds to the molar enthalpy of $X$= liq, hex. The findings of these free energy calculations are given in the SI.

\newpage

\begin{table*}
	\sffamily
	\begin{sansmath}
	\begin{center}
		\caption{\label{tab:spec}\textbf{Technical specifications of the MD simulations and the order parameter.} For the LJ system, all quantitates are in the LJ dimensionless units.}
		\begin{tabular}{lccc}
			\hline\hline
			~~~~~~& ~~~~TIP4P/Ice~~~~ & ~~~~mW~~~~ & ~~~~Lennard-Jones~~~~ \\
			\hline
			time step & 2~fs & 2~fs & 0.00002--0.0025 \\
			thermostat time constant & 200~fs & 200~fs & 0.25\\
			barostat time constant & 2~ps & 2~ps & 2.5 \\
			Distance cutoff, $r_c$ & 3.2~\AA & 3.2~\AA & 1.40 \\
			Type of $\textbf{q}_6$ & Regular & Regular & neighbor-averaged \\
			$q_{6,c}$ & 0.5 & 0.5 & 0.3 \\
			\hline
		\end{tabular}
	\end{center}
	\end{sansmath}
\end{table*}

\newpage

\begin{figure*}[hcp]
	\begin{sansmath}
	\includegraphics[width=0.8713\textwidth]{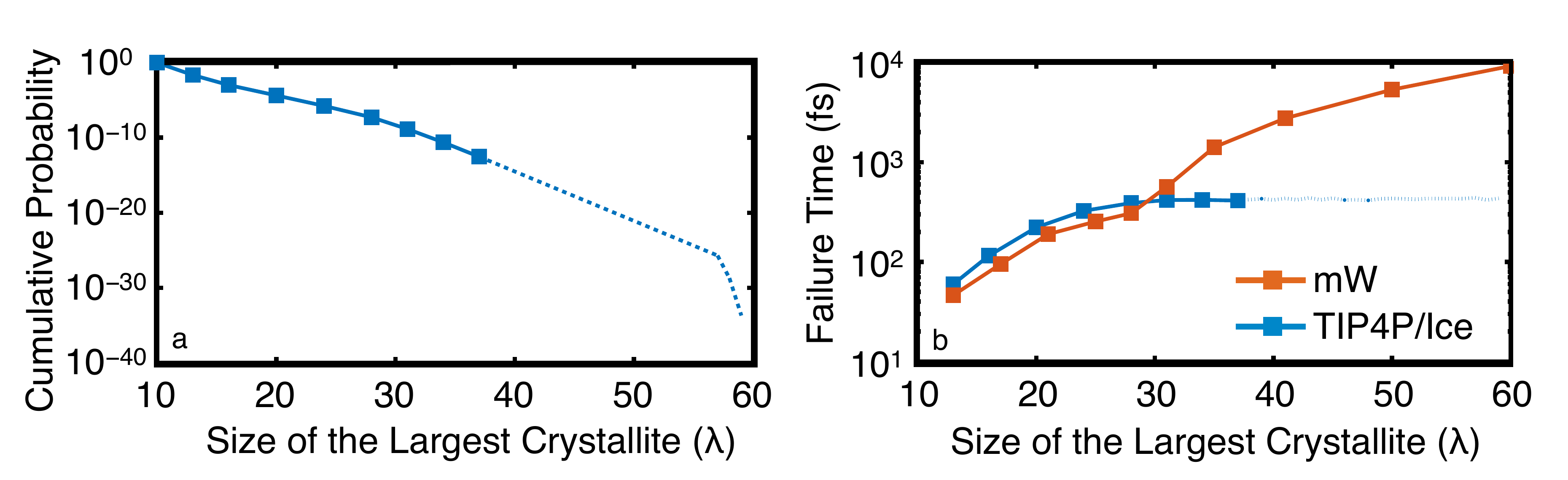}
	\caption{\label{fig:ffs:unsuccessful}
	\textbf{The failure of the conventional FFS approach in the TIP4P/Ice system at 230~K and 1~bar.} All symbols are obtained from actual simulations, while the dashed lines are schematic representations of what would happen upon performing more FFS iterations.  (a) $P(\lambda|\lambda_1)$ vs.~$\lambda$ does not have the positive curvature observed in successful FFS calculations presented in the main Figs.~1,~\ref{fig:lj-low}a and~\ref{fig:mW}a. (b) Average failure times for trajectories aimed at $\lambda$. Beyond $\lambda\approx30$, this average failure time plateaus. This suggests that the addition of new water molecules to the largest crystallites is only nominal and does not lead to a meaningful improvement in the overall structural quality of the arising configurations. We observe a strong correlation between the plateauing of the average failure time and the failure of the corresponding FFS calculation, and based on this heuristic, we terminate the calculation depicted in (a) at $\lambda\approx40$. Contrast this to the strictly increasing average failure time in the successful FFS calculation in the mW system.
	}
	\end{sansmath}
\end{figure*}

\newpage

\begin{figure*}[hcp]
	\centering
	\includegraphics[width=.7398\textwidth]{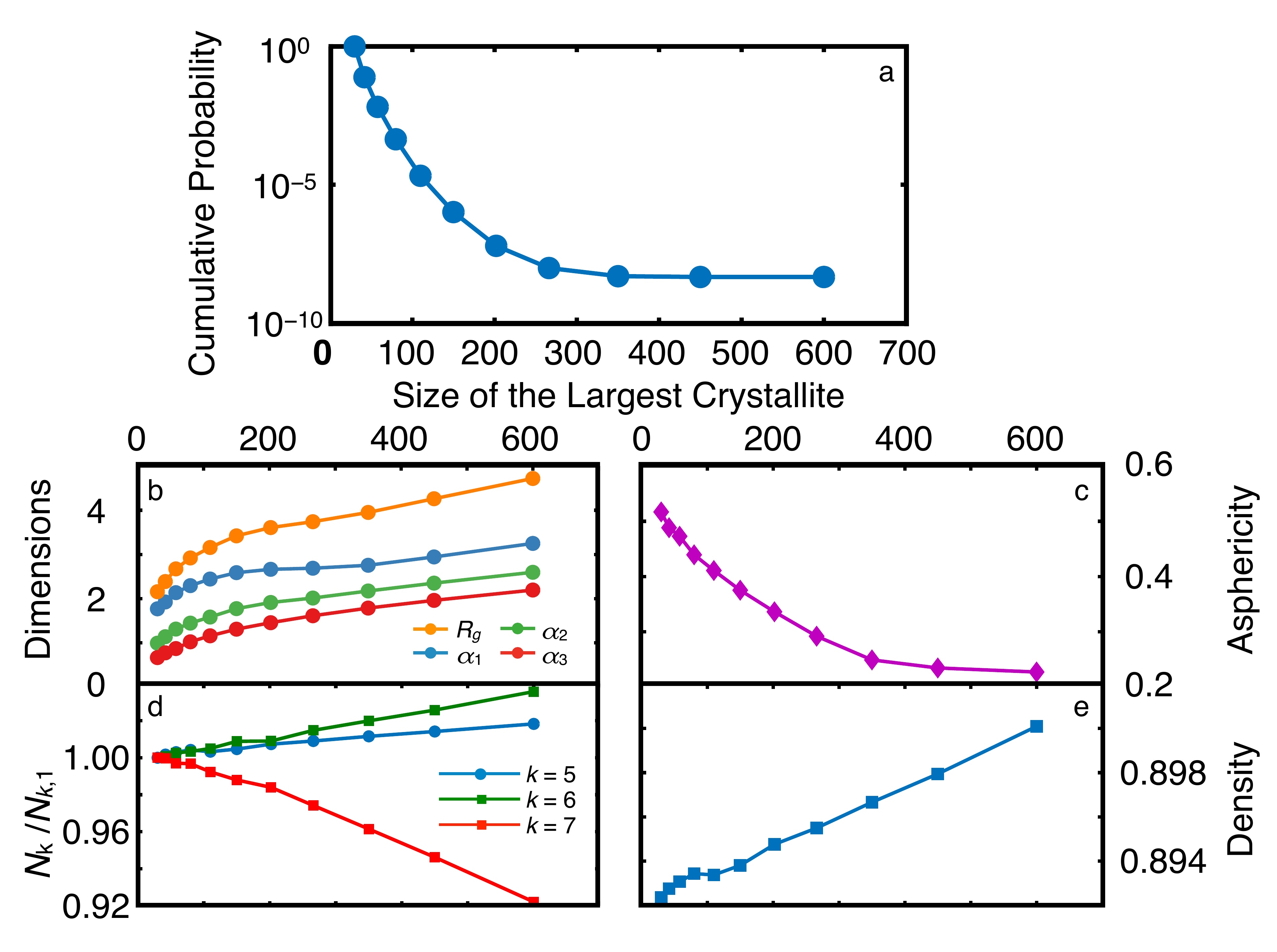}
	\caption{\label{fig:lj-low}
	\textbf{Crystallization of the LJ system close to the triple point.} FFS calculations are performed at $k_BT/\epsilon=0.48$ and $p\sigma^3/\epsilon=0$. (a) No inflection is observed in the cumulative probability curve. Furthermore, (b) the dimensions and (c) the asphericity of the largest crystallite, (d) the number of three-, four- and five-member rings and (e) the density of the system change monotonically between the liquid and the crystal. The observed lack of inflection and non-monotonicity in the calculations presented here reveal that the trends presented in the insets of Figs.1 and 4 are also observed in low-pressure LJ systems.
	}
\end{figure*}

\newpage

\begin{figure*}[hcp]
	\centering
	\includegraphics[width=.7405\textwidth]{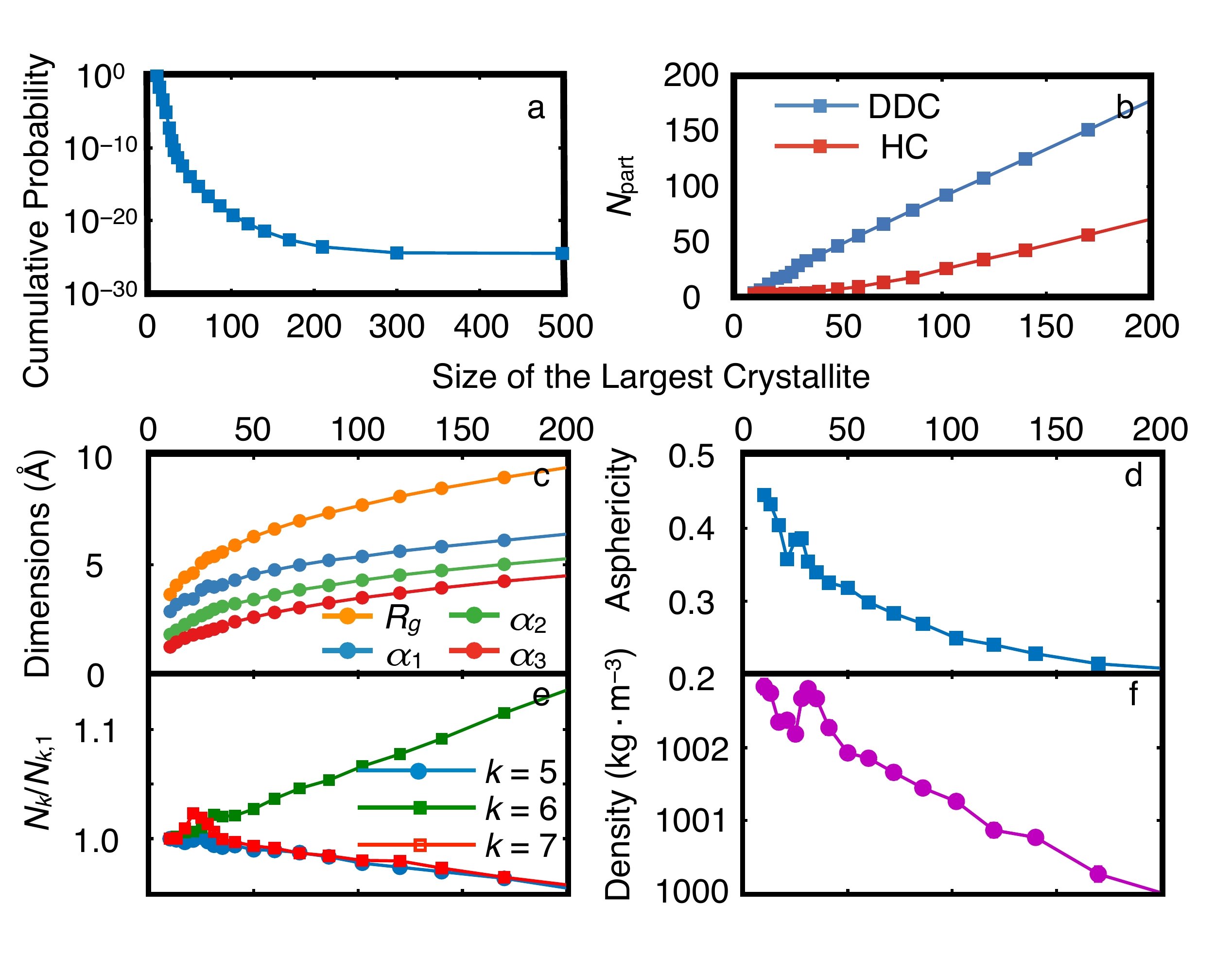}
	\caption{\label{fig:mW}
	\textbf{Ice nucleation in the mW system at 230~K and 1~bar.} (a) Cumulative probability, (b) cage participation, (c) shape and (d) asphericity of the largest crystallite, (e) number of five-, six- and seven-member rings and (f) density as a function of the size of the largest crystallite. Note that the inflection in cumulative probability, and the associated non-monotonicities in density, asphericity and ring statistics are very mild in the mW system, and no monotonicity exists in the dimensions and the radius of gyration of the largest crystallite.
	}
\end{figure*}

\newpage

\begin{figure*}
	\centering
	\includegraphics[width=0.8465\textwidth]{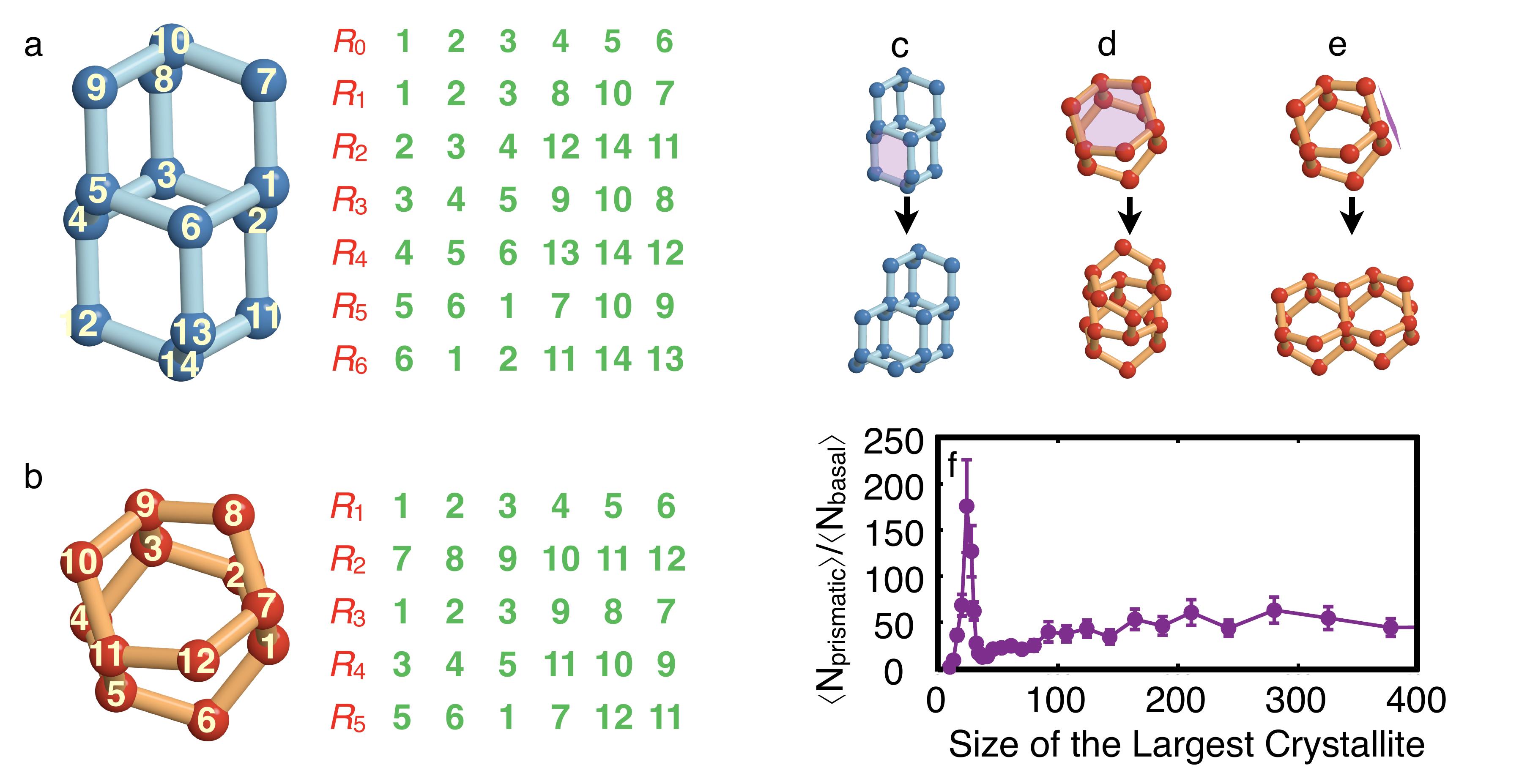}
	\caption{\label{fig:dd}
	\textbf{Topological features and growth characteristics of different cages}. (a) Topological features of a double-diamond cage. Every DDC has one equatorial ring $R_0$, and six peripheral rings $R_1,\cdots,R_6$. Every water molecule in $R_0$ participates in four hexagonal rings. For instance, $5$ participates in $R_3,R_4$ and $R_5$ in addition to $R_0$. Every triplet along $R_0$ is crossed by \emph{exactly} one other ring in the DDC. For instance, the triplet $(1,2,3)$ is crossed by $R_1$. The three top peripheral rings, $R_1,R_3$ and $R_5$, and the three bottom peripheral rings, $R_2,R_4$ and $R_6$, each have one water molecule in common, namely $10$ and $14$, respectively. (b) Topological features of a hexagonal cage. $R_1$ and $R_2$ are the basal planes of the cage, while $R_3,R_4$ and $R_5$ are the prismatic planes. These are not real two-dimensional planes due to their bending as a result of tetrahedral arrangement of hydrogen bonds. (c-e) Schematic representation of the available pathways for the formation of new DDCs and HCs. (c) Each DDC has six identical six-member rings that can act as anchoring points for new DDCs or HCs. (d-e) Each HC has two distinct sets of six-member rings as anchoring points for new cages. The basal plane (d) of a hexagonal cage can support the attrition of both HCs and DDCs. The prismatic plane of an HC (e), however, only supports the attrition of new HCs. There are far fewer basal connections in the system as depicted in (f). 
	}
\end{figure*}

\newpage

\begin{figure*}[hcp]
	\centering
	\includegraphics[width=0.5195\textwidth]{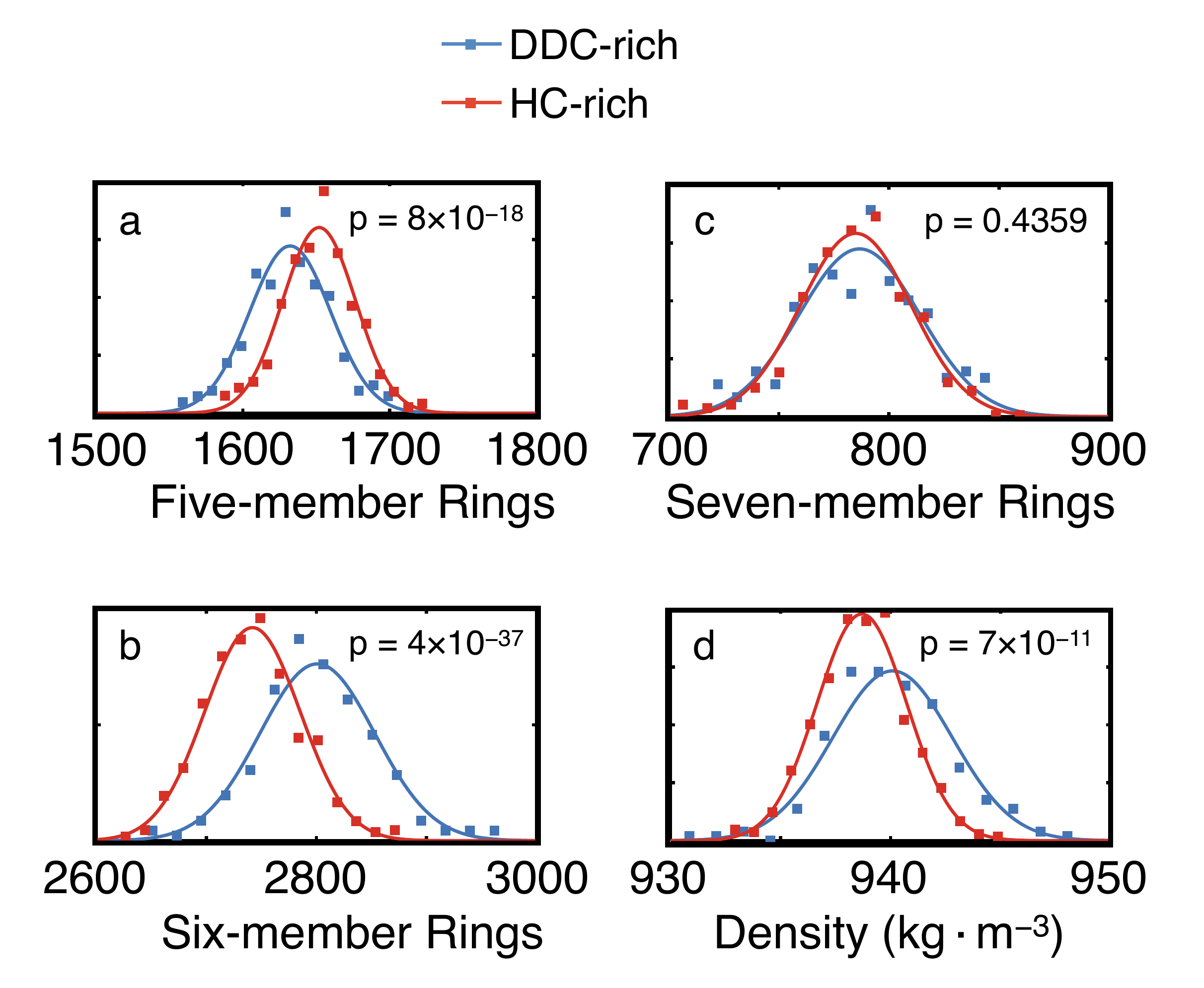}
	\caption{\label{fig:dr_nonmonot}
	\textbf{Non-monotonicities in ring statistics and density.} Distribution of ring populations (a-c) and densities (d) in configurations that are rich in DDCs (blue) and rich in HCs (red). In each panel, $p$ is the probability that these distinct distributions are statistically indistinguishable, and is computed from student's $t$-test analysis. In order to better visualize these distributions, a Gaussian with the same mean and standard deviation is plotted for every distribution. DDC- and HC-rich configurations are distinguished using the k-mean clustering algorithm.  
	}
\end{figure*}

\begin{figure*}[hcp]
	\centering
	\includegraphics[width=0.9488\textwidth]{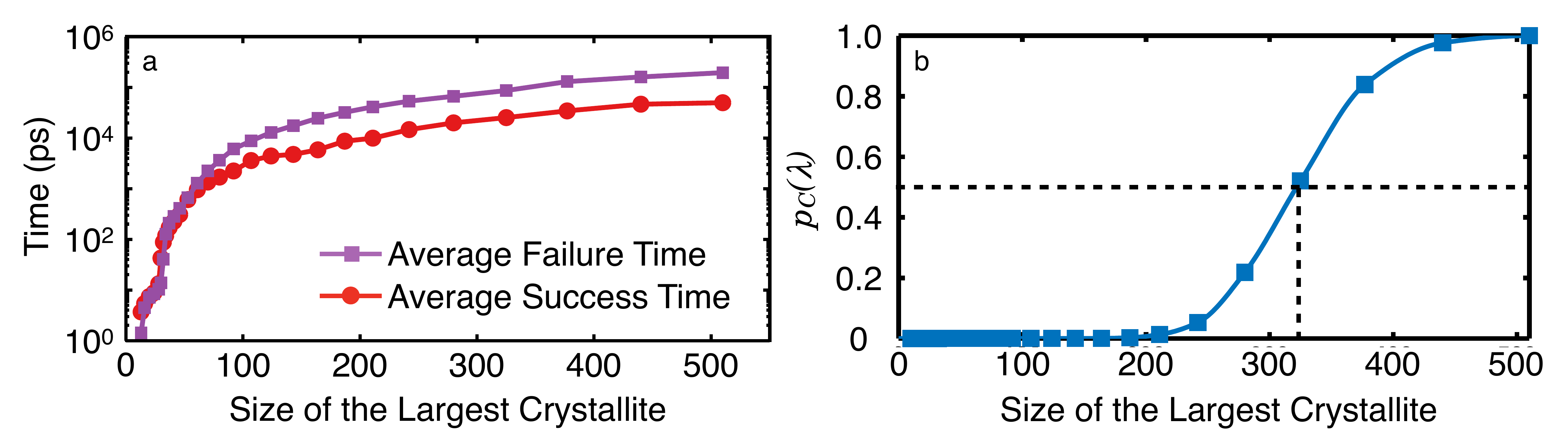}
	\caption{\label{fig:trajTime}\textbf{Computational cost and the approximate commitor probability} (a) Average success and failure times for the trajectories initiated at different iterations of our FFS calculation in the TIP4P/Ice system. (b) $p_C(\lambda)$ vs.~$\lambda$ for the FFS calculation of the nucleation rate in the TIP4P/Ice system. The critical nucleus has  $320\pm20$ water molecules.  }
\end{figure*}

\end{document}